\documentclass{jpsj-suppl}
\usepackage{txfonts} 
\usepackage{url}
\usepackage{graphicx}

\title{Production Uncertainties of p-Nuclei in the $\gamma$-Process in Massive Stars
Using a Monte Carlo Approach}

\author{T. \textsc{Rauscher}$^{1,2,7}$, N. \textsc{Nishimura}$^{3,7}$, R. \textsc{Hirschi}$^{3,4,7}$, G. \textsc{Cescutti}$^{2,7}$, A. St.J. \textsc{Murphy}$^{5,7}$, and A. \textsc{Heger}$^{6}$}

\inst{$^{1}$Department of Physics, University of Basel, Klingelbergstr.\ 82, 4056 Basel, Switzerland \\
$^{2}$Centre for Astrophysical Research, Univ. of Hertfordshire, College Lane, Hatfield AL10 9AB, UK \\
$^{3}$Astrophysics group, Lennard-Jones Laboratories, Keele University, ST5 5BG, Staffordshire, UK\\
$^{4}$Kavli Institute for the Physics and Mathematics of the Universe (WPI),
University of Tokyo, 5-1-5 Kashiwanoha, Kashiwa, 277-8583, Japan\\
$^{5}$SUPA, School of Physics and Astronomy, University of Edinburgh, Edinburgh EH9 3FD, UK\\
$^{6}$School of Physics and Astronomy, Clayton Campus, Monash University, Victoria 3800, Australia\\
$^{7}$UK Network for Bridging Disciplines of Galactic Chemical Evolution (BRIDGCE), UK, \url{www.bridgce.net}
}

\recdate{August 15, 2016}

\abst{Proton-rich nuclei, the so-called p-nuclei, are made in photodisintegration processes
in outer shells of massive stars in the course of the final supernova explosion. Nuclear uncertainties in the production of these nuclei have been quantified in a Monte Carlo procedure. Bespoke temperature-dependent uncertainties were assigned to different types of reactions involving nuclei from Fe to Bi and all rates were varied randomly within the uncertainties. The resulting total production uncertainties of the p-nuclei are below a factor of two, with few exceptions. Key reactions dominating the final uncertainties have been identified in an automated procedure using correlations between rate and abundance uncertainties. Our results are compared to those of a previous study manually varying reaction rates.}

\begin{document}
\maketitle

\section{Introduction}

A number of proton-rich nuclei between Se and Hg, the so-called p-nuclei, cannot be produced in the s- and r-processes. The majority of p-nuclei are made in photodisintegration processes
in outer shells of massive stars in the course of the final supernova explosion, the $\gamma$-process. Whether this is sufficient
to explain the terrestrial p-abundances remains an open question, as this is the result of a superposition of nuclei produced in various sources over time \cite{arngor,p-review}.

Predictions of the p-nucleus production suffer from several types of uncertainties, from the chosen site \cite{trav}, over numerical treatments \cite{rau15,rau16}, to uncertainties in the input. Here, we focus on the astrophysical reaction rates as the source of uncertainty in final abundances.  We present
large-scale Monte Carlo (MC) variations in the full $\gamma$-process reaction network, using temperature-dependent rate uncertainties combining experimental and theoretical uncertainties.
From detailed statistical analyses, realistic uncertainties in the final p-abundances
are derived. Furthermore, based on rate and abundance correlations an automated
procedure is able to identify the most important reactions in complex flow patterns from
superposition of many zones. This method is superior to visual
inspection of flows and manual variation of limited rate sets. The resulting list of
important (and uncertain) reactions is a valuable resource for experimentalists and
theorists seeking to improve abundance predictions.

\section{Rate variation}

\begin{figure}[tbh]
\begin{center}
\includegraphics[width=0.53\columnwidth]{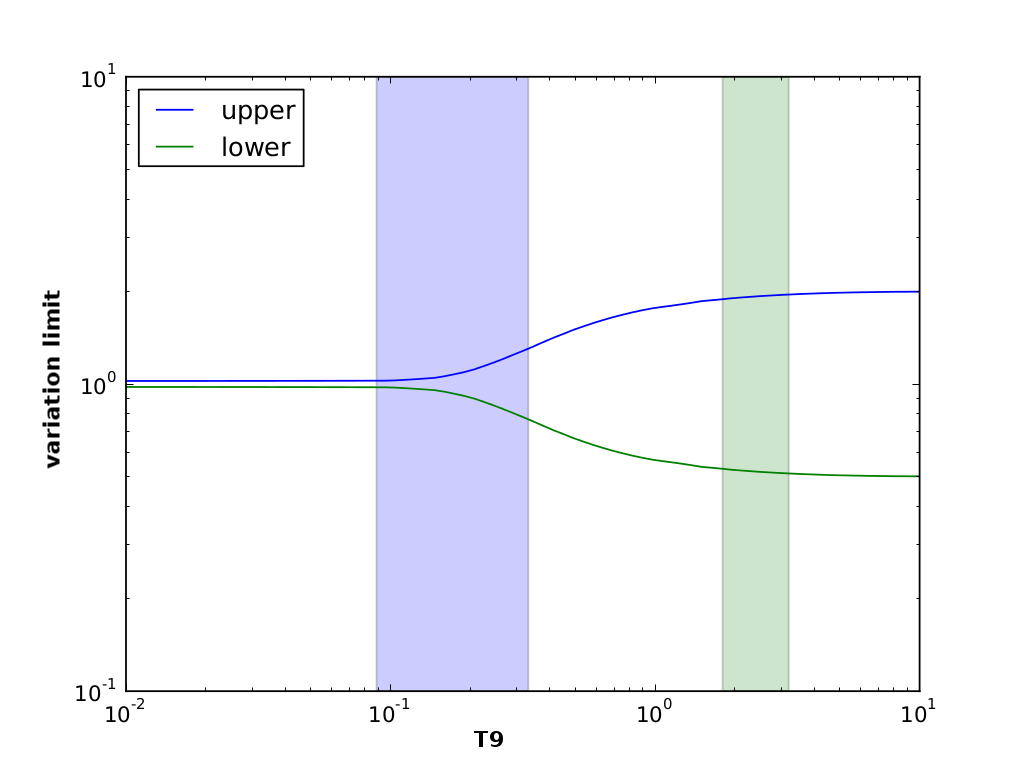}
\includegraphics[width=0.46\columnwidth]{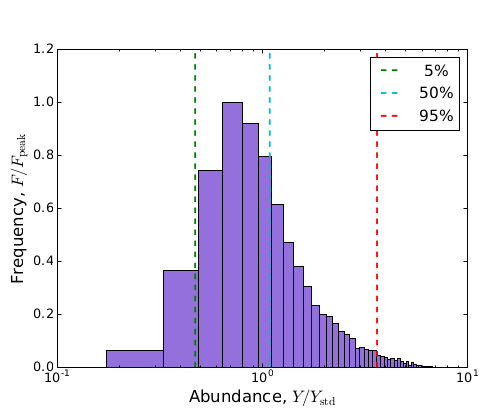}
\end{center}
\caption{Left panel: Example of the temperature-dependent variation range. Shown are the upper and lower limits of the variation factor of the rate for $^{157}$Gd(n,$\gamma$)$^{158}$Gd as function of plasma temperature $T_9$ in 10$^9$ K. Shaded areas indicate the relevant $T_9$ ranges for the s-process (left) and the $\gamma$-process (right).\\
Right panel: Example of the abundance distributions obtained in the MC runs. The bounds encompassing 5\% and 95\% of the distribution are marked to allow to use them as uncertainty measures.\label{fig:varfact}}
\end{figure}

Due to elevated temperatures in the supernova shock passage and high level density of the nuclei involved, experiments only partially constrain rates because they only measure reaction cross sections of nuclei in their ground states (g.s.). With increasing plasma temperature $T$, more and more nuclei are thermally excited and reactions on excited states increasingly contribute to the stellar rate. Just as the stellar rate is given by a superposition of reactions on g.s.\ and excited states, also the uncertainty has to be constructed as such a combination. Using the g.s.\ contribution $X_0(T)$ to the stellar rate, the total uncertainty factor $u^*(T)$ of a reaction rate can be given as \cite{stellarerrors,advances}
\begin{equation}
u^*(T)=U_{\mathrm{exp}}+\left( U_\mathrm{th}-U_{\mathrm{exp}} \right) \left[ 1-X_0(T)\right] \label{eq:uncertainty} \quad,
\end{equation}
where $U_{\mathrm{exp}}$ and $U_\mathrm{th}$ are the experimental and theoretical uncertainty factors, respectively, and with $U_\mathrm{th}>U_{\mathrm{exp}}$. Assuming a symmetric uncertainty, this would limit the range of rate variation factors to $u^*$ and $1/u^*$. An example for the $T$ dependence of these limits is shown in Fig.\ \ref{fig:varfact} (left) for the reaction rate of $^{157}$Gd(n,$\gamma$)$^{158}$Gd. Although the ground-state cross section is tightly constrained experimentally \cite{bao}, reactions on excited states of $^{157}$Gd contribute significantly to the stellar rate at increased temperatures. In this case, theoretical uncertainties start to become important already at s-process temperatures and they dominate at typical $\gamma$-process temperatures.

In our MC calculations different uncertainty limits were assigned to different reaction types, with the temperature dependence obtained from Eq.\ (\ref{eq:uncertainty}). Experimental uncertainties $U_{\mathrm{exp}}$ were considered for g.s.\ contributions when available, taken from \cite{p-review,kadonis,jina}. Theoretical uncertainties $U_\mathrm{th}$ for g.s.\ and excited state contributions were assigned symmetric or asymmetric uncertainties, as appropriate, which are assumed to include systematic errors. In particular, predicted rates for neutron-induced reactions received an uncertainty limit of a factor of 2 (0.5), whereas an asymmetric uncertainty was used for predicted rates involving protons (factor of 2.0 (0.33)) and $\alpha$ particles (factor 2.0 (0.1)). The same variation factor is used for forward and reverse rate.
The MC variation factors provided by the random number generator are values $0\leq f_\mathrm{MC}\leq 1$, drawn from a uniform distribution. The actual varied rate $r(T)$ is computed from
\begin{equation}
r(T)=r_{\mathrm{lo}}(T)+f_\mathrm{MC}\left( r_{\mathrm{hi}}(T)-r_{\mathrm{lo}}(T)\right) \quad.\label{eq:rate}
\end{equation}
The upper and lower rate limits $r_{\mathrm{hi}}$ and $r_{\mathrm{lo}}$ are determined by the uncertainty limits given above. Note that they depend on $T$ and can be asymmetric. It is also important to note that $f_\mathrm{MC}$ does not depend on $T$ as otherwise this would result in non-analytic rates.

Figure \ref{fig:varfact} (right) shows an example of the obtained uncertainty distribution in a final abundance, resulting from the combined uncertainties of all contributing rates. The more reactions are contributing, the closer the distribution shape will be to a lognormal distribution.

\section{Results and Comparison to Previous Study}

\begin{figure}[tbh]
\begin{center}
\includegraphics[width=\columnwidth]{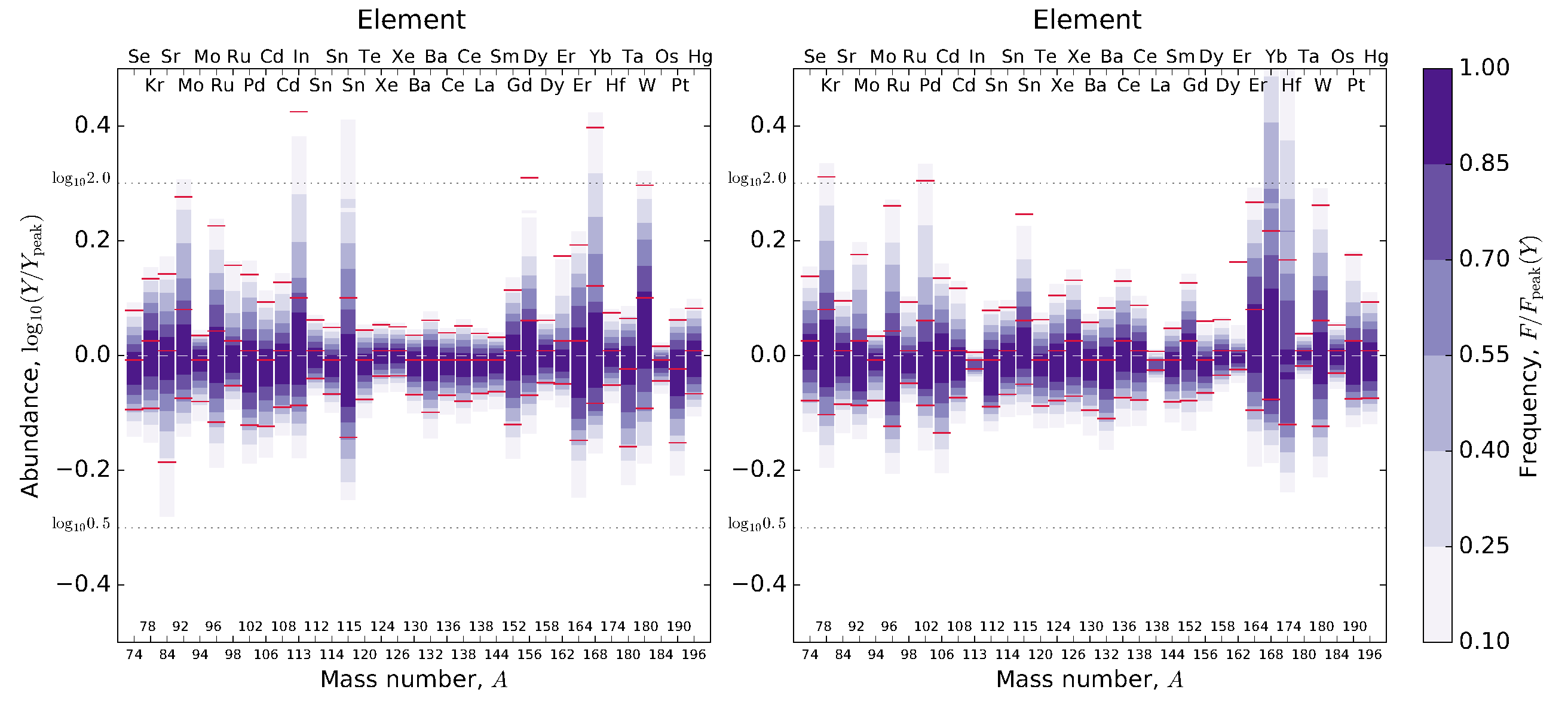}
\end{center}
\caption{Total production uncertainties of the classical p-nuclides in the explosion of a 15 $M_\odot$ (left) and a 25 $M_\odot$ (right) solar metallicity star, obtained with trajectories from \cite{rhhw}. The color shade is the probabilistic frequency and the 90\% probability intervals marked for each nuclide (see Fig.\ \ref{fig:varfact} (right panel) for further details). Horizontal dashed lines indicate a factor of two uncertainties.\label{fig:final}}
\end{figure}

The full MC study ran 10000 rate variations in about 100 trajectories with a network containing 3800 nuclides, for several stellar models. Figure \ref{fig:final} shows the final production uncertainties for a 15 $M_\odot$ and a 25 $M_\odot$ solar metallicity star \cite{rhhw}. Note that the shown uncertainties are a weighted average from the uncertainties arising in each trajectory, i.e., in each zone of the stellar model. Most uncertainties are below a factor of two, with a few exceptions. The largest uncertainties are found for p-nuclei whose production is not limited to a single zone but rather is spread out over many zones \cite{rau15,rau16}. This leads to the situation that several reactions may contribute to the production or destruction of the p-nucleus due to the different peak temperature encountered in each zone.

It is possible for several p-nuclides, nevertheless, to identify ``key reactions'' which dominate the production uncertainty of the given p-nucleus. When such key reactions are better constrained theoretically or experimentally (where possible), then the related uncertainty will be reduced considerably. Due to the complexity of the flow pattern in the $\gamma$-process and the combined action of many reactions, it is not easily possible to manually identify such key reactions. In the analysis of our MC data, we make use of a correlation procedure which finds key reactions by the strong correlation of rate variations with final production uncertainties (see \cite{rau16} for details). As very uncertain rates may cover other rates' uncertainties, we distinguish several levels of key reactions. Level 1 key reactions are the most important ones, level 2 reactions can affect the remaining uncertainties once level 1 reactions have been determined, and so on. The full lists of level $1-3$ key reactions for the 15 $M_\odot$ and 25 $M_\odot$ stars are given in \cite{rau16}. It was shown that an improved knowledge of comparatively few reactions may reduce the production uncertainties considerably. Unfortunately, they are mostly reactions with unstable target nuclei or exhibit strong excited-state contributions the stellar rate, which limits the feasibility of an experimental determination.

Most such key reactions are of the type (n,$\gamma$)$\leftrightarrow$($\gamma$,n), plus a few (p,$\gamma$)$\leftrightarrow$($\gamma$,p) and ($\alpha$,$\gamma$)$\leftrightarrow$($\gamma$,$\alpha$). It is interesting to compare these to the reactions assigned to be important in the study of \cite{rapp}. In this study, sets of reaction rates were manually varied in postprocessing of 14 zones of a 25 $M_\odot$ stellar model different to ours. Their key reactions are largely different to our findings.

Tables 2 and 3 in \cite{rapp} showed reactions with the strongest impact on p-abundances in their approach. We cannot confirm the importance of (n,p) reactions, as no (n,p) reactions were identified as key reactions in our approach. Among the remaining (p,$\gamma$)$\leftrightarrow$($\gamma$,p) rates, we only confirm $^{91}$Nb + p $\leftrightarrow$ $\gamma$ + $^{92}$Mo and $^{77}$Br + p $\leftrightarrow$ $\gamma$ + $^{78}$Kr as level 1 key rates. These were marked as ``particularly important'' in \cite{rapp}. Another ``particularly important'' rate, $^{95}$Tc + p $\leftrightarrow$ $\gamma$ + $^{96}$Ru, appears as level 2 rate, only after $^{92}$Mo + $\alpha$ $\leftrightarrow$ $\gamma$ + $^{96}$Ru has been determined, which is our level 1 key rate for the production of $^{96}$Ru. This rate is also the only one of the ($\alpha$,$\gamma$)$\leftrightarrow$($\gamma$,$\alpha$) rates in Table 2 of \cite{rapp} (also marked as ``particularly important'') we confirm. We also do not find full ($\gamma$,$\alpha$) chains among the key rates but our level 1 key rates $^{160}$Er + $\alpha$ $\leftrightarrow$ $\gamma$ + $^{164}$Yb and $^{176}$W + $\alpha$ $\leftrightarrow$ $\gamma$ + $^{180}$Os appear in two of the chains pointed out by \cite{rapp}. On the other hand, we find additional key reactions of all types not listed in \cite{rapp}, see \cite{rau16} for details.

The reason for the differing results is threefold. On one hand, the zoning of their stellar model is much cruder and it appears that not all inner zones possibly contributing to the production of p-nuclei have been taken into account \cite{rau15,rau16}. This omits certain peak temperatures and finer ranges of peak temperatures and thus assigns a different importance to flow paths and related reactions. On the other hand, the approach of manually varying rates has two shortcomings. The previous study varied rates by somewhat arbitrarily chosen factors without considering experimental uncertainties or a $T$ dependence of the uncertainty. A rate with a large uncertainty can have more impact on the final production uncertainty than a tightly constrained rate, even when the abundance is more sensitive to a variation of the latter. Moreover, variations of single rates cannot account for the combined action of multiple rate variations where uncertainties may enhance or cancel each other. This is even true for the variation of groups of reactions applying the same multiplication factor, as performed in \cite{rapp}.

Only a MC approach varying individual rates simultaneously by different factors is able to include both effects and thus is superior to a simple variation of isolated rates or groups of rates.

\end{document}